\documentclass{iesmart}               

\usepackage{graphicx}          
\newcommand{\SetR}{{\rm I\!R }}

\newcommand{\bx}  { {\bf x} }

\newcommand{\by}  { {\bf y} }

\begin{document}
\begin{frontmatter}

\title{
A DC programming approach for the constrained two-dimensional
non-guillotine cutting problem\thanksref{footnoteinfo}} 
\thanks[footnoteinfo]{Corresponding author Mahdi Moeini. Tel. +33-141131388. Fax +33-141131272.}

\author[Paestum]{Mahdi Moeini}\ead{mahdi.moeini@ecp.fr},    
\author[Rome]{Hoai An Le Thi}\ead{lethi@univ-metz.fr},               

\address[Paestum]{Laboratoire G\'enie Industriel (LGI), \'Ecole Centrale Paris, Grande Voie des Vignes, F-92 295, Ch\^{a}tenay-Malabry, France.}  
\address[Rome]{Laboratoire d'Informatique
Th\'eorique et Appliqu\'ee
 (LITA), Universit\'e Paul Verlaine - Metz,\\ Ile du Saulcy, 57045, Metz, France.}             

\begin{abstract}
We investigate a new application of DC (Difference of Convex
functions) programming and DCA (DC Algorithm) in solving the
constrained two-dimensional non-guillotine cutting problem. This
problem consists of cutting a number of rectangular pieces from a
large rectangular object. The cuts are done under some constraints
and the objective is to maximize the total value of the pieces
cut. We reformulate this problem as a DC program and solve it by
DCA. The performance of the approach is compared with the standard
solver CPLEX.
\end{abstract}
\end{frontmatter}

\section{Introduction}
\label{sec:Introduction}

The field of combinatorial optimization involves many challenging
problems. Different practical applications of these problems,
motivates the researchers to develop new methods in order to solve
them as efficiently as possible. One of the important classes of
combinatorial optimization problems is the class of the cutting
and packing problems. The constrained two-dimensional
non-guillotine cutting problem (NGC) is one of the cutting and
packing problems that have been studied by several researchers
(\cite{Beasley1985,Beasley2004,Nepomuceno,Wang}). The constrained
two-dimensional non-guillotine cutting problem consists of cutting
a number of rectangular pieces from a large rectangular object.
The cuts are done under some constraints and the objective is to
maximize the total value of the pieces cut.

\begin{figure*}
\hrule
\centering{\includegraphics[height=5cm]{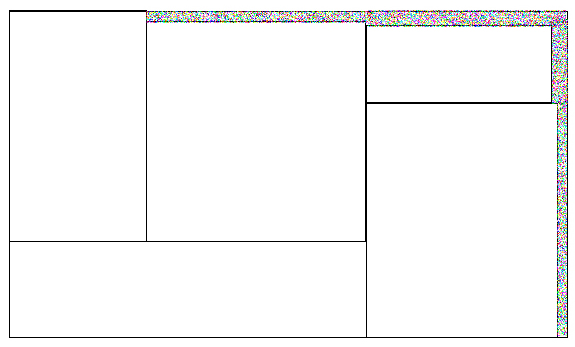}}%
\hrule%
\caption{A two-dimensional non-guillotine cutting pattern in which
the produced wastes are minimized} \label{Fig:Waste}
\end{figure*}

This problem arises in several practical applications, such as
cutting the steel or glass plates into required sizes, cutting the
wood sheets to make furniture etc. \cite{Beasley1985}. The optimal
solution of this problem minimizes the amount of wastes produced
(see Fig. \ref{Fig:Waste}).

In this study, we consider the constrained two-dimensional
non-guillotine cutting problem and we investigate a deterministic
approach based on DC programming techniques to solve it. The
particular interest of the work is in the design of the algorithms
called DCA (DC Algorithm). This approach is a local deterministic
method based on DC (Difference of Convex functions) programming.
The DC Algorithm (DCA) was first introduced, in its preliminary
form, by Pham Dinh Tao in 1985 and has been extensively developed
since 1994 by Le Thi Hoai An and Pham Dinh Tao. It becomes now
classic and popular. DCA has been successfully applied to many
large-scale (smooth or nonsmooth) non-convex programs in various
domains of applied sciences, for example \emph{tomography}
\cite{WSSH04}, \emph{finance}
\cite{Moeini2009a,Moeini2009b,Moeini2010}, and \emph{machine
learning} \cite{Liu:jcgs05,neuman,Ronan:icml06}. Numerical
experiments show that DCA is in many cases more robust and
efficient than standard methods (see e.g.
\cite{Harring05,lethithesis,lethi2005,Moeini2009a,Moeini2009b,Liu:jcgs05,Moeini2010,neuman,PLT97,PLT98,WSSH04}
and reference therein).

In this work, we first formulate the underlying optimization
models in the form of a DC programs in which a DC function is
minimized over a closed convex set. Then, DCA is used to solve it.
Computational experiences, performed over standard benchmark
problems, show that DCA is quite efficient for solving the
constrained two-dimensional non-guillotine cutting problem and
compares favorably with the standard solver ILOG CPLEX.

The structure of the paper is as follows. The constrained
two-dimensional non-guillotine cutting problem is reviewed in the
section \ref{sec:NGCreview}. In section \ref{sec:GeneralDCetDCA}
we give an outline of general DC programs and DCA. The DC program
for the constrained two-dimensional non-guillotine cutting problem
is discussed in section \ref{sec:DCandDCAforNGC}. The
computational experiences are reported in section
\ref{sec:ComputExperim} and the last section includes some
conclusions.

\section{The constrained two-dimensional non-guillotine cutting problem}
\label{sec:NGCreview}

Consider a large rectangular object. In the constrained
two-dimensional non-guillotine cutting problem, we are interested
in cutting some smaller rectangular pieces from this large
rectangular object. We suppose that the cuts are done in a way
that the edges of the smaller pieces are parallel to the edges of
the large rectangular object and there must be no overlapping
between the pieces cut. Furthermore, we suppose that the
orientation of each of the smaller objects is known in advance and
a limited number of the small pieces is available. There is a
positive integer number associated to each rectangle that
indicates its value. The objective of the constrained
two-dimensional non-guillotine cutting problem is to cut the
smaller pieces from the large rectangle in a way that the total
value of the pieces cut be maximized.

The constrained two-dimensional non-guillotine cutting problem is
known to be an NP-Complete problem, so it is is very difficult to
solve the problem efficiently. The problem has been the object of
several articles. The articles \cite{Beasley1985},
\cite{Beasley2004}, and \cite{Nepomuceno} summarize some of the
works done on this problem.

\subsection{Mathematical Formulation}%
\label{sec:MathFormulationNGC}

In this section we present the mathematical formulation of the
constrained two-dimensional non-guillotine cutting problem. We use
the notations used by Napoleao et al. \cite{Nepomuceno}. This
formulation concerns a $0-1$ linear programming model that has
been already presented in \cite{Beasley1985}. There is also a
nonlinear programming formulation that has been introduced in
\cite{Beasley2004}.

Suppose that $m$ types of pieces are available. For each type $i$
of the pieces, we know the characteristics of the object, such as
its length and width $(l_i,w_i)$, its value $v_i$ and we know that
only a limited number of the piece $i$ is available, which is
noted by $b_i$. The pieces must be cut from a large rectangular
object with the length $L$ and the width $W$. We use a binary
variable $x_{ipq}$ to say whether or not the piece $i$ can be cut
orthogonally from the large object at the position $(p, q)$:
$$
 \begin{array}{ll}
 x_{ipq}  =  \left\{
 \begin{array}{ll}
 1,\mbox{if a piece of type $i$ is allocated at position $(p, q)$,} \\
 0,\mbox{otherwise.} \\
\end{array}
\right.%
\end{array}
$$
One can assume that $p$ and $q$ belong, respectively, to the
following sets (\cite{Beasley1985,Nepomuceno}):%
\begin{equation}
P:=\left\{p :p=\sum_{i=1}^{m}\alpha_{i}l_{i}, p\leq L -
\min\{l_{i},i=1,\dots,m\}, \alpha_{i}\geq 0, \alpha_{i} \in
\mathbf{Z}\right\},
\end{equation}%

\begin{equation}
Q:=\left\{q :q=\sum_{i=1}^{m}\beta_{i}w_{i}, q\leq W -
\min\{w_{i},i=1,\dots,m\}, \beta_{i}\geq 0, \beta_{i} \in
\mathbf{Z}\right\}.
\end{equation}%

The cuts cannot pass over the edges of the large rectangle, so we
need to take into account the following sets:
\begin{equation}
P_{i}:=\left\{p :p \in P, p\leq L - l_{i}\right\},
\end{equation}%

\begin{equation}
Q_{i}:=\left\{q :q \in Q, q\leq W - w_{i}\right\}.
\end{equation}%

\begin{figure*}
\hrule
\centering{\includegraphics[height=5cm]{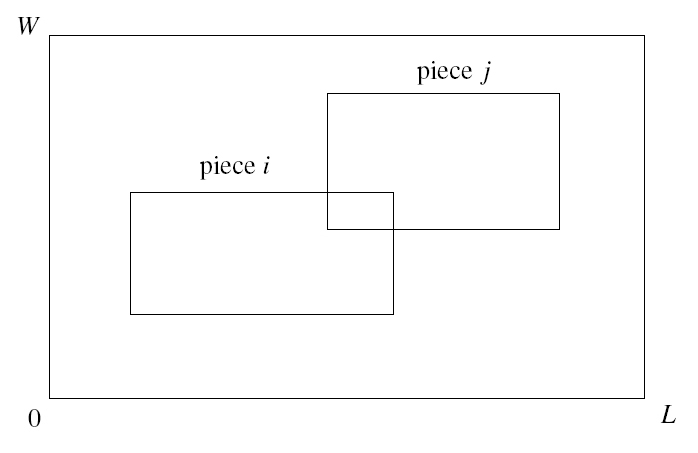}}%
\hrule \caption{Overlapping of the pieces $i$ and $j$.}
\label{Fig:Overlapping}
\end{figure*}

The coefficients $a_{ipqrs}$ are defined as follows in order to
prevent the interposition of the pieces cut (see Fig.
\ref{Fig:Overlapping}):
$$
 \begin{array}{ll}
 a_{ipqrs}  =  \left\{
 \begin{array}{ll}
 1,\mbox{if $p\leq r \leq p+l_{i}-1$ and $q\leq s \leq q+w_{i}-1$,} \\
 0,\mbox{otherwise.} \\
\end{array}
\right.%
\end{array}
$$
The $a_{ipqrs}$ are defined for each piece $i=1,\dots,m$, and for
each coordinates $(p,q)$ as well as $(r,s)$.

Using these notations, we can now present the mathematical
formulation of the constrained two-dimensional non-guillotine
cutting problem:
\begin{eqnarray}
\nonumber (NGC): \hspace{60mm} \\
\nonumber \max \; \sum_{i=1}^{m}\sum_{p \in P_i}\sum_{q \in Q_i}v_i x_{ipq} \hspace{20mm} \\
\nonumber s.t.\hspace{45mm}\\
\nonumber \sum_{i=1}^{m}\sum_{p \in P_i}\sum_{q \in
 Q_i}a_{ipqrs} x_{ipq} \leq 1 & & : \forall r\in P,s\in Q, \hspace{10mm}\\
\nonumber \sum_{p \in P_i}\sum_{q \in
 Q_i}x_{ipq} \leq b_i & & :i=1,\dots,m,  \\
\nonumber x_{ipq} \in \{0,1\} & & :\forall i,p, q.
\end{eqnarray}

By solving this problem, one maximizes the total value associated
to the all pieces cut from the large rectangular object. \\
The model includes $(|P| |Q| + m)$ constraints and $n:=
(\sum_{i=1}^{m}\sum_{p \in P_i}\sum_{q \in Q_i} 1)$ variables. It
means that the problem takes a huge dimension even by moderate
values of $|P_i|$ or $|Q_i|$ ($i=1,\dots,m$).

\section{General DC programs and DCA}%
\label{sec:GeneralDCetDCA}%

Let $\Gamma _{0}(\mathrm{I\!R}^{n})$ denotes the convex cone of
all lower semi-continuous proper convex functions on
$\mathrm{I\!R}^{n}$. Consider the following primal DC program
$$
(P_{dc})  \;\;\;  \quad \beta_p = \inf\{F(x):=g(x)-h(x) \;: \;
x\in \mathrm{I\!R}^{n}\},
$$
where $g, h\in \Gamma _{0}(\mathrm{I\!R}^{n})$.

A DC program $(P_{dc})$ is called a DC polyhedral program when
either $g$ or $h$ is a polyhedral convex  function (i.e., the
pointwise supremum of a finite collection of affine functions).
Note that a polyhedral convex function is almost always
differentiable, say, it is differentiable everywhere except on a
set of measure zero.

Let $C$ be a nonempty closed convex set. Then, the problem
\begin{equation}
\inf\{f(x):=g(x)-h(x) \; : \; x\in C\},
\end{equation}
can be transformed into an unconstrained DC program by using the
indicator function of $C$, i.e.,
\begin{equation}
\inf\{f(x):=\phi(x)-h(x)\; : \;x\in \mathrm{I\!R}^{n}\},
\end{equation}
where $\phi:=g+\chi _{C}$ is   in $\Gamma
_{0}(\mathrm{I\!R}^{n})$.

Let $g^{\ast}(y):= \sup\{\langle x,y \rangle-g(x) : x\in
\mathrm{I\!R}^{n}\}$ be the conjugate function of $g$. Then, the
following program is called the dual program of ($P_{dc}$):
\begin{equation}
(D_{dc}) \;\;\;\;\;  \quad \beta_d = \inf\{h^{\ast} (y)-g^{\ast}
(y)\; : \;y\in \mathrm{I\!R}^{n}\}.
\end{equation}
Under the natural convention in DC programming that is $+ \infty
-(+\infty)= + \infty $,
and by using the fact that every function $%
h\in $ $\Gamma _{0}(\mathrm{I\!R}^{n})$ is characterized as a
pointwise supremum of a collection of affine functions, say
\[
h(x):=\sup \{\langle x,y\rangle -h^{\ast }(y) \; : \; y\in
\mathrm{I\!R}^{n}\},
\]
it can be proved that $\beta_p = \beta_d$ \cite{PLT98}. There is a
perfect symmetry between primal and dual DC programs, that is the
dual of $(D_{dc})$ is $(P_{dc})$.

Recall that, for $\theta \in \Gamma _{0}(\mathrm{I\!R}^{n})$ and $%
x_{0}\in dom$ $\theta :=\{x\in \mathrm{I\!R}^{n} | \theta
(x_{0})<+\infty \}$,  the subdifferential of $\theta $ at $x_{0}$,
denoted $\partial \theta (x_{0})$,  is  defined as
\begin{equation}
\partial \theta (x_{0}):=\{y\in \mathrm{I\!R}^{n}: \theta (x)\geq \theta
(x_{0})+\langle x-x_{0},y\rangle ,\forall x\in \mathrm{I\!R}^{n}\}
\label{subdif}
\end{equation}
which is a closed convex set in $\mathrm{I\!R}^{n}$. It
generalizes the derivative in the sense that $\theta $ is
differentiable at $x_{0}$ if and only if $\partial \theta (x_{0})$
is reduced to a singleton which is exactly $\{\nabla
\theta(x_{0})\}$.

The necessary local optimality condition for the primal DC
program, $(P_{dc})$, is
\begin{equation}
\label{local}
\partial h(x^\ast) \subset \partial g(x^\ast).
\end{equation}
The condition (\ref{local}) is also sufficient for many important
classes of DC programs, for example, for DC polyhedral programs,
or when function $f$ is locally convex at $x^{\ast }$
\cite{lethi2005,PLT97}.

A point $x^*$ satisfies the generalized Kuhn-Tucker condition
\begin{equation}
\label{eqcr}
\partial h(x^\ast)
\cap
\partial g(x^\ast)\neq \emptyset
\end{equation}
 is called a critical point of
$g-h$. It follows that if $h$ is a polyhedral convex function,
then a critical point
 of $g-h$ is almost always a local solution to $(P_{dc})$.

The transportation of global solutions between $(P_{dc})$ and
$(D_{dc})$ is expressed by
\cite{lethithesis,lethi2005,PLT97,PLT98}:

\noindent {\bf Property 1:}
\begin{equation}
 \label{eqiii}
\lbrack \cup _{y^{\ast }\in \mathcal{D}}\,\partial g^{\ast
}(y^{\ast })]\subset \mathcal{P}, \quad [\cup _{x^{\ast }\in
\mathcal{P}}\,\partial h(x^{\ast })]\subset \mathcal{D},
\end{equation}
where\ $\mathcal{P}$ and $\mathcal{D}$ denote the solution sets of
$(P_{dc})$ and $(D_{dc})$ respectively.

Under certain technical conditions, this property also holds  for
the local solutions of $(P_{dc})$ and $(D_{dc})$
\cite{lethithesis,lethi2005,PLT97,PLT98}. For example the
following result holds:

\noindent {\bf Property 2:} Let $x^*$ be a local solution to
$(P_{dc})$ and let $y^* \in \partial h(x^*)$. If $g^*$ is
differentiable at $y^*$ then $y^*$ is a local solution to
$(D_{dc})$. Similarly, let $y^*$ be a local solution to $(D_{dc})$
and let $x^* \in \partial g^*(y^*)$. If $h$ is differentiable at
$x^*$ then $x^*$ is a local solution to $(P_{dc})$.

Based on local optimality conditions and duality in DC
programming, the DC Algorithm (DCA) consists in   constructing two
sequences $\{x^{l}\}$ and $\{y^{l}\}$  of trial solutions for the
primal and dual programs, respectively, such that the sequences
$\{g(x^{l})-h(x^{l})\}$ and $\{h^{\ast}(y^{l})-g^{\ast }(y^{l})\}\
$ are decreasing, and
 $\{x^{l}\}$ (resp.$\{y^{l}\}$) converges  to a primal  feasible solutions $\widetilde{x}$
(resp. a dual feasible solution $\widetilde{y}$)
  satisfying   the local optimality condition  and
\begin{equation}
\widetilde{x}\in \partial g^{\ast }(\widetilde{y}),\quad
\widetilde{y}\in
\partial h(\widetilde{x}).  \label{intersubdif}
\end{equation}

DCA then yields the next simple scheme:
\begin{equation}
y^{l}\in \partial h(x^{l});\quad x^{l+1}\in \partial g^{\ast
}(y^{l}). \label{DCAscheme}
\end{equation}
In other words, these two sequences $\{x^{l}\}$\ and $\{y^{l}\}$\
are determined in the way that $x^{l+1}$  and   $y^{l+1}$ are
solutions of the convex primal program $(P_{l})$ and dual program
$(D_{l+1})$, respectively. These are defined as
\begin{eqnarray}
(P_l)  \;\;\;  &  \inf  \{ g(x)-h(x^{l})-\langle x-x^{l},y^{l}\rangle \; : \; x\in \mathrm{I\!R}^{n} \}, & \\
(D_{l+1})  \;\;\;  &  \inf  \{h^{\ast }(y)-g^{\ast
}(y^{l})-\langle y-y^{l},x^{l+1}\rangle \; : \;y\in
\mathrm{I\!R}^{n} \}. &
\end{eqnarray}
At each iteration, the DCA performs a double linearization with
the use of the subgradients of $h$ and $g^{\ast }$. In fact, in
each iteration, one replaces in the primal DC program, $(P_{dc})$,
the second component $h$ by its affine minorization
$h_{l}(x):=h(x^{l})+\langle x-x^{l},y^{l}\rangle $ to construct
the convex program $(P_{l})$ whose solution set is nothing but
$\partial g^{\ast }(y^{l})$. Likewise, the second DC~component
$g^{\ast }$ of the dual DC program, $(D_{dc})$, is replaced by its
affine minorization $g^{\ast }_{l}(y):=g^{\ast }(y^{l})+\langle
y-y^{l},x^{l+1}\rangle $   to obtain the convex program
$(D_{l+1})$ whose $\partial h(x^{l+1})$ is the solution set. Hence
DCA works with the convex DC components $g$ and $h$ but not with
the DC function $f$ itself. Moreover, a DC function $f$ has
infinitely many DC decompositions which have crucial impacts on
the performance of the DCA in terms of speed of convergence,
robustness, efficiency, and globality of computed solutions.
Convergence properties of the DCA
 and its theoretical basis are described in
\cite{lethithesis,lethi2005,PLT97,PLT98}. However, it is
worthwhile to summarize the following properties for the sake of
completeness :
\begin{itemize}
\item DCA is a descent method (\emph{without line search}).   The
sequences\ $\{g(x^{l})-h(x^{l})\}$\ and\ $\{h^{\ast
}(y^{l})-g^{\ast }(y^{l})\}$\ are\ decreasing such that
$$g(x^{l+1})-h(x^{l+1})\leq h^{\ast }(y^{l})-g^{\ast }(y^{l}) \leq
g(x^{l})-h(x^{l}).$$
\item If $g(x^{l+1})-h(x^{l+1})=g(x^{l})-h(x^{l})$, then $x^{l}$
is a critical point of $g-h$ and $y^{l}$ is a critical point of
$h^{\ast }-g^{\ast }$. In this case,  DCA terminates at $l^{th}$
iteration. \newline
\item If the optimal value $\beta_p$ of problem $(P_{dc})$ is
finite and the infinite sequences $\{x^{l}\}\ $and $\{y^{l}\}$\
are bounded, then every limit point $\widetilde{x}\ $(resp.
$\widetilde{y}$) of the
sequence $\{x^{l}\}$\ (resp. $\{y^{l}\})$\ is a critical point of $g-h$%
\ (resp. $h^{\ast }-g^{\ast }$). \newline
\item DCA has linear convergence for general DC programs. For
polyhedral DC programs the sequences $\{x^{l}\}$ and $\{y^{l}\}$
contain finitely many elements and the algorithm converges to a
solution in a finite number of iterations.

\end{itemize}

\section{DC program for the constrained
two-dimensional non-guillotine cutting problem}%
\label{sec:DCandDCAforNGC}%

We consider a new approach based on DC programming and DCA to
solve the constrained two-dimensional non-guillotine cutting
problem. The DCA requires a reformulation of the problem so that
the objective function be represented by the difference of two
convex functions. Then the original problem becomes a DC program
in which the DC function is minimized over a convex set. In this
section, we introduce the corresponding DC reformulations of the
NGC problem and then present a DCA to solve the corresponding DC
program.

Using the exact penalty result presented in \cite{lethi2005b}, we
will formulate (NGC) in the form of a DC minimization problem with
linear constraints which is consequently a DC program. Let\\
$\begin{array}{cc}
 {A} := &  \{\bx \in
[0,1]^{n}:\sum_{i=1}^{m}\sum_{p \in P_i}\sum_{q \in
 Q_i}a_{ipqrs} x_{ipq} \leq 1 : \forall r\in P,s\in Q,\sum_{p \in P_i}\sum_{q \in
 Q_i}x_{ipq} \leq b_i  :i=1,\dots,m\}
\end{array}
$.

Let $\alpha(\bx)$ be the concave function defined as follows
\[\alpha(\bx):=\sum_{i=1}^{m}\sum_{p \in P_i}\sum_{q \in
 Q_i} x_{ipq}(1 - x_{ipq}),\]%

The concave function $\alpha(x)$ is non-negative on $A$ hence
(NGC) can be re-written as follows
 \[\mbox{(NGC - 2) : }\quad min \left\{- \sum_{i=1}^{m}\sum_{p \in P_i}\sum_{q \in Q_i}v_i x_{ipq} : \alpha(\bx) \leq 0, \bx \in A\right\}.\]

Since the objective function is linear, $A$ is a bounded
polyhedral convex set, and the concave function $\alpha(x)$ is
non-negative on $A$; according to \cite{lethi2005b}, there is
$t_{0}\geq 0$ such that for any $t> t_{0}$, the program (NGC - 2)
is equivalent to
\[\mbox{(NGC - 3) : }\quad min \left\{F(\bx) :=- \sum_{i=1}^{m}\sum_{p \in P_i}\sum_{q \in Q_i}v_i x_{ipq} + t \alpha(\bx) : \bx \in A\right\}.\]

The function $F$ is concave in variables $\bx$; consequently it is
a DC function. A natural DC formulation of the problem (NGC - 3)
is
\[\mbox{(NGC-DC) : }\quad min \left\{F(\bx) := g(\bx) - h(\bx): \bx \in \SetR^{n} \right\},\]
 where%
 \[g(\bx) = - \sum_{i=1}^{m}\sum_{p \in P_i}\sum_{q \in Q_i}v_i
 x_{ipq} + \chi_{A}(\bx)\]
 and
 \[h(\bx) = t \sum_{i=1}^{m}\sum_{p \in P_i}\sum_{q \in
 Q_i} x_{ipq}(x_{ipq} - 1).\]%
Here $\chi _{A}$ is the indicator function on $A$, i.e. $\chi
_{A}(\bx)=0$ if $(\bx)\in A$ and $+\infty$ otherwise.

\subsection{DCA for solving (NGC-DC)}
According to the general framework of DCA, we first need computing
a sub-gradient of the function $h(\bx)$ defined by $h(\bx) = t
\sum_{i=1}^{m}\sum_{p \in P_i}\sum_{q \in
 Q_i} x_{ipq}(x_{ipq} - 1)$.
 From the definition of $h(\bx)$ we have

\begin{equation}
\by^{k} \in \partial h(\bx^{k})\Leftrightarrow y_{ipq}^{k}:= t (2
x_{ipq}^{k} - 1),
\end{equation}
for $i=1,\dots,m$, $p \in P_i$, and $q \in Q_i$.

Secondly, we need to compute an optimal solution of the following
linear program
\begin{equation}
\min \left\{- \sum_{i=1}^{m}\sum_{p \in P_i}\sum_{q \in Q_i}v_i
 x_{ipq} - \langle \bx,\by^{k}\rangle :\bx \in A \right\}
\end{equation}
that will be $\bx^{k+1}$. To sum up, the DCA applied to (NGC-DC)
can be described as follows.%

\noindent \textbf{Algorithm DCA}
\begin{enumerate}
    \item \textbf{Initialization}: Choose $\bx^{0}\in \SetR^{n}$, $\epsilon>0$, $t > 0$, and set
    $k=0$.%
    \item \textbf{Iteration}:\\
    Set $y_{ipq}^{k}:= t (2 x_{ipq}^{k} - 1)$ for $i=1,\dots,m$, $p \in P_i$, and $q \in Q_i$.\\
    Solve the following linear program
    \[
    \min \left\{- \sum_{i=1}^{m}\sum_{p \in P_i}\sum_{q \in Q_i}v_i
    x_{ipq} - \langle \bx,\by^{k}\rangle :\bx \in A \right\}
    \]
    to obtain $\bx^{k+1}$.\\
    \item If $\left\Vert \bx^{k+1}- \bx^{k}\right\Vert \leq \epsilon$ then \textbf{STOP} and take $\bx^{k+1}$ as an optimal
    solution, otherwise set
    $k=k+1$ and go to step 2.
\end{enumerate}

\noindent{\bf Finding a good initial point for DCA}

In fact, one of the key questions in DCA is how to find a good
initial solution for it. In this work, we solved a relaxed form of
the (NGC-DC) program to find a good initial solution. To this aim,
we relax the following constraints from the (NGC-DC) program:
\begin{equation}
\label{R-S-Constraints}%
\sum_{i=1}^{m}\sum_{p \in P_i}\sum_{q \in Q_i}a_{ipqrs}
x_{ipq}\leq 1 \quad : \quad \forall r\in P,s\in Q,%
\end{equation}
 and we add them to the objective function of the (NGC-DC) program. In order to penalize any
 violation of the relaxed constraints, a sufficiently large positive real number $u$ is used as the penalty
 parameter.

 Let $\beta(\bx)$ be the function defined by
\[%
\beta(\bx):=max_{r\in P,s\in Q} \left(0, (\sum_{i=1}^{m}\sum_{p \in P_i}\sum_{q \in Q_i}a_{ipqrs} x_{ipq})-1\right),%
\]%
and define the convex set $B$ as follows
\[B=\left\{\bx \in \SetR^{n}: \sum_{p \in P_i}\sum_{q \in
 Q_i}x_{ipq} \leq b_i, 0 \leq x_{ipq} \leq 1, i=1,\dots,m\right\}.\]

Using the function $\beta(\bx)$ and the parameter $u$ for relaxing
the constraints (\ref{R-S-Constraints}) we obtain the following
program
\begin{equation}
\label{NGC-Initial-1}%
min \left\{G(\bx) := - \sum_{i=1}^{m}\sum_{p \in P_i}\sum_{q \in
Q_i}v_i x_{ipq} + t \alpha(\bx) + u \beta(\bx) : \bx \in B
\right\}.
\end{equation}
$G(\bx)$ is a DC function and a DC formulation of
(\ref{NGC-Initial-1}) can be

\begin{equation}
\label{NGC-Initial-2}%
min \left\{G(\bx) := \varphi(\bx) - \phi(\bx): \bx \in \SetR^{n} \right\},%
\end{equation}
 where%
 \[\varphi(\bx) = - \sum_{i=1}^{m}\sum_{p \in P_i}\sum_{q \in Q_i}v_i
 x_{ipq} + \chi_{B}(\bx)\]
 and
 \[\phi(\bx) = t \sum_{i=1}^{m}\sum_{p \in P_i}\sum_{q \in
 Q_i} x_{ipq}(x_{ipq} - 1) - u . max_{r,s} \left(0, (\sum_{i=1}^{m}\sum_{p \in P_i}\sum_{q \in Q_i}a_{ipqrs} x_{ipq})-1\right).\]%
Here $\chi _{B}$ is the indicator function on $B$.

The solution of (\ref{NGC-Initial-2}) is used as the initial point
for \textbf{Algorithm DCA}. The solution of (\ref{NGC-Initial-2})
may not be feasible to (NGC-DC), but we need just one iteration of
DCA to obtain a feasible solution of (NGC-DC) and all the other
iterations of DCA will improve the solution.

In fact, we have tested DCA from different initial points, some of
them are:
\begin{itemize}
\item The point obtained by the above procedure;%
\item $\bx = (0,\dots,0)$;%
\item $\bx = (1,\dots,1)$;%
\item The optimal solution of the relaxed (NGC) problem obtained
by replacing the binary constraints $x_{ipq}\in \{0,1\}$ by $0
\leq x_{ipq} \leq 1$ for all $i,p,q$.
\end{itemize}
According to our experiments the initial point provided by the
first procedure is the best.

Since (\ref{NGC-Initial-2}) is a DC program, we use again DCA for
solving it. The DCA applied on this problem is described as
follows:

\noindent \textbf{Algorithm Initial-DCA}
\begin{enumerate}
    \item \textbf{Initialization}: Choose $\bx^{0}\in \SetR^{n}$, $\epsilon>0$, $t > 0$, $u > 0$, and set $k=0$;
    \item \textbf{Iteration}:\\
    Set
    $$
    \begin{array}{ll}
    y_{ipq}^{k}:=\left\{
    \begin{array}{ll}
    t (2 x_{ipq}^{k} - 1), \mbox{if all $(r,s)$ constraints are satisfied at the point } \bx^{k}, \\
    t (2 x_{ipq}^{k} - 1) - u (a_{ipqrs}), \mbox{otherwise: for some ($r$, $s$)},\\
    \end{array}
    \right.%
    \end{array} \nonumber
    $$%
    for $i=1,\dots,m$, $p \in P_i$, and $q \in Q_i$.\\
    Solve the following linear program to obtain $\bx^{k+1}$:
    \[
    \min \{- \sum_{i=1}^{m}\sum_{p \in P_i}\sum_{q \in Q_i}v_i
    x_{ipq} - \langle \bx,\by^{k}\rangle :\bx \in B \}
    \]
    \item If $\left\Vert \bx^{k+1}- \bx^{k}\right\Vert \leq \epsilon$ then \textbf{STOP} and take $\bx^{k+1}$ as an optimal
    solution to (\ref{NGC-Initial-2}), otherwise set
    $k=k+1$ and go to step 2.
\end{enumerate}

Clearly we need an initial point to start \textbf{Algorithm
Initial-DCA}. This time we use the solution of the linear program
obtained by relaxation of the binary constraints in (NGC) (the
binary constraints $x_{ipq} \in \{0,1\}$ are replaced by $0\leq
x_{ipq} \leq 1$ for all $i,p,q$).

\section{Computational experiences}%
\label{sec:ComputExperim}%

Some experiments have been carried out to evaluate the quality of
the solutions provided by the proposed algorithm. The solutions
have been compared with the results given by the standard solver
CPLEX version 11.2. This solver have been used to solve the binary
programming (NGC) problem.

The experiments have been performed over 12 benchmark test
problems. The test problems, taken from literature
\cite{Beasley1985,Beasley2004,Nepomuceno,Wang}, are also
available through:\\%
$http://people.brunel.ac.uk/\sim mastjjb/jeb/orlib/ngcutinfo.html$.%

The algorithms are coded in C++ and run on
 a Pentium IV $3.0$ $GHz$ with $1GB$ $RAM$. The precision $\varepsilon$ has been set equal to $10^{-6}$ for
DCA and a CPU time limit of $60$ seconds has been considered for
the solver CPLEX. We used the solver CPLEX version 11.2 in order
to solve the linear sub-problems generated at each iteration of
DCA.

\begin{table}[htbp]
  \caption{Results of the experiments carried out over 12 benchmark test problems.}
  \label{tab:Results}
  \begin{tabular*}{\hsize}{llllllllllll}
\hline
       &      &          &       &        &  CPLEX &           &  DCA   &           &        &      &  \\\hline%
       & (L;W)& $m$-$|P|$-$|Q|$  & N. Var. & N. Con. & CPU  &  Opt. val.& CPU    &  Opt. val.&  iter. & $t$  &  $u$   \\
\hline
1   &  (10;10)  &   5 -  8 -  6  &    60      &     51        &   0.140   &   164  &   0.062   &   146   &   4  &    30   &    10     \\ 
2   &  (10;10)  &   7 - 10 - 10  &     250     &     107       & 0.891   &   230  &   0.156   &   212   &   5  &    30   &    30     \\ 
3   &  (10;10)  &  10 - 10 - 10  &     411     &     110       &   0.281   &   247  &   0.156   &   242   &   4  &    20   &     5    \\ 
4   &  (15;10)  &   5 -  3 - 10  &     68     &     35        &  0.031   &   268  &   0.063   &   268   &   4  &    25   &    25    \\ 
5   &  (15;10)  &   7 - 10 - 10  &     154     &     107       & 0.047   &   358  &   0.329   &   358   &  23  &    20   &    80     \\ 
6   &  (15;10)  &  10 - 13 - 10  &     552     &     140       & 12.094   &   289  &   1.019   &   283   &  26  &    25   &   200     \\ 
7   &  (20;20)  &   5 - 20 - 20  &     763     &     405       &  0.110   &   430  &   0.328   &   404   &   4  &    50   &   100     \\ 
8   &  (20;20)  &   7 -  6 - 20  &     343     &     127       &  9.719   &   834  &   0.234   &   828   &   6  &    50   &   100     \\ 
9   &  (20;20)  &  10 - 20 - 18  &     1413     &     370       &  3.094   &   924  &   2.531   &   924   &   8  &     5   &     5    \\ 
10  &  (30;30)  &   5 - 30 - 26  &     363     &     785       &  4.860   &  1452  &   0.562   &  1452   &   4  &   100   &   100      \\ 
11  &  (30;30)  &   7 - 21 - 27  &     1120     &     574       & 60.266   &  1688  &   1.094   &  1688   &   4  &   100   &   100      \\ 
12  &  (70;40)  &  20 - 45 - 18  &     3657     &     830       & 60.312   &  2726  &   7.204   &  2568   &   4  &   180   &    10      \\ \hline%
\end{tabular*}
\end{table}

The results are presented in Table \ref{tab:Results}. The table
contains some information about the test problems: the length
($L$) and the width ($W$) of the large rectangular object, the
number of the types of the pieces to be cut and the number of the
elements in the sets $P$ and $Q$. The table contains also some
information about the (NGC) problems corresponding to each test
problem: number of the variables (N. Var.) and number of the
constraints (N. Con.). In this table, the CPU time (CPU) in
second, the best optimal value (Opt. val.) of the solver CPLEX and
also DCA, the number of DCA iterations (iter.), and the values of
the penalty parameters ($t$ and $u$) are presented.

The proposed approach has presented very satisfactory results in
comparison to the solver CPLEX. The solver CPLEX solves
efficiently
 some first test problems that are small size problems, but CPLEX needs more time to solve the others.
Specially, we consider the test problems number $6$ and $8-12$.
For these problems, DCA has a better performance. The results are
particularly interesting for the test problems number $10$ and
$11$, for which the proposed DCA method gives the same solutions
as CPLEX, but in a significantly shorter CPU time.

\section{Conclusion}%
\label{sec:Conclusion}%

In this paper, we present a new approach based on DC programming
and DCA to solve the constrained two-dimensional non-guillotine
cutting problem. We saw that DCA outperforms in some cases the
commercial solver CPLEX.

The computational results suggest to us extending the numerical
experiments in higher dimensions, and combining DCA and
Branch-and-Bound algorithms for globally solving the constrained
two-dimensional non-guillotine cutting problem. Work in these
directions is currently in progress.


\end{document}